# A Root Extension to the Propagator Algorithm for Direction of Arrival Estimation

Pradeep Dheerendra[1], Zoltan Derzsi[2]

1 – Biomedical Engineering Division, James Watt School of Engineering, University of Glasgow, Glasgow, UK

2 - Center for Artificial Intelligence and Robotics, New York University Abu Dhabi, UAE

*Abstract*—In this paper, we propose an extension to the Propagator algorithm for source bearing estimation by performing root decomposition which eliminates the need for spectral search over angles. Further the propagator spatial spectrum is reused after performing the polynomial rooting to alleviate the observed drawback of an increase in RMSE at lower SNRs for estimating the directions-of-arrival of signals impinging on a linear array of sensors. Our proposed algorithm achieves a 98% reduction in computational complexity compared to Propagator alongside an improved angular resolution suitable for real time DOA estimation for wireless communications.

*Keywords- Direction of Arrival (DOA) estimation; source bearing estimation; propagator method; root-MUSIC*

## I. INTRODUCTION

The direction-of-arrival (DOA) or source bearing estimation [1-5] is the method of estimating the directions (angles) from which a signal is impinging on an array of sensors. One of the most popular subspace-based DOA methods is Multiple signal classification (MUSIC) algorithm [6, 7]. This is a subspace method based on exploiting the eigen structure of the input covariance matrix. Thus, it involves eigen value decomposition (EVD), eigen vector computation of the cross-spectral matrix (CSM) and a spectral search over angles to estimate both the number of incident signals, and the DOA of each. This results in a computationally intensive approach which limits its implementation in real-time systems.

The Root-MUSIC algorithm [8] achieves a complexity reduction over MUSIC by performing polynomial rooting to find DOAs instead of spectral search over angles. This also achieves an improved threshold estimation performance [9] compared to MUSIC. When proposed it was applicable only to uniformly spaced linear arrays (ULA), later methods like [10-16] have extended it to non-ULA configurations as well. While Root-MUSIC reduces complexity, it still suffers from the complexity as EVD is still required, which is too high when the number of sensors is large.

The Propagator Method (PM) proposed in [17, 18] achieves a reduction in complexity by eliminating the need for EVD and eigen vectors. This is achieved by a linear operator called 'propagator' which only depends on the array steering vectors and can be estimated from CSM using a least squares process. The computational gain of propagator over MUSIC is of the order of ratio of number of sources to number of sensors [19, 20], hence especially advantageous when the number of sensors is too large. Despite several improvements to Propagator [21-26], the spectral search over all angles is not eliminated, and hence the complexity has not been fully reduced. Thus, it presents further scope for reduction in complexity of Propagator.

This paper is structured as follows. First the array signal model used is enunciated. Next, the method of polynomial rooting for estimating the DOA is applied to propagator method eliminating the need for performing a spectral search over all angles. This proposed root-propagator method eliminates both the EVD and the spectral scan over angles when compared to traditional MUSIC algorithm enabling a huge reduction in complexity. This also results in an improved angular resolution, however an increase in root mean squared error (RMSE) at lower SNRs is also observed.

In the later part, a further modification to root-propagator method is proposed to overcome this drawback, by reusing the propagator spatial spectrum after performing root decomposition to achieve a good performance-to-complexity tradeoff. Finally, the simulation results are presented to support the claims. It is followed by an analysis on computational complexities of all the algorithms discussed.

The terminologies used in the paper are as follows: Superscripts $(\cdot)^T$ and $(\cdot)^H$ denote Transpose and Hermitian transpose respectively and upper/lower case bold letters denote matrix/vector.

## II. ARRAY SIGNAL MODEL

Consider an array composed of M omni-directional sensors located in x-y plane and assume that D (D < M) narrow-band signals, centered on a known frequency, say $\omega_c$, impinges on the array from distinct unknown directions $\theta_1$, $\theta_2$ ... $\theta_D$ with respect to the normal of the array. For simplicity assuming the sources and sensors are located in the same plane and the sources are in the far field of the array. In this case, the only parameter that characterizes source location is its DOA ($\theta$). Here in this paper, only the estimation of the azimuth angle $\theta$ is considered.

Using complex envelope representation, the M×1 vector received by the array can be expressed, in matrix notation, as

$$\boldsymbol{u}(t) = \boldsymbol{A}(\theta) \cdot \boldsymbol{s}(t) + \boldsymbol{n}(t)$$

where,

$$\boldsymbol{u}(t) = [u_1(t)\, u_2(t)\, \ldots\, u_M(t)]^T$$
$$\boldsymbol{s}(t) = [s_1(t)\, s_2(t)\, \ldots\, s_D(t)]^T$$
$$\boldsymbol{n}(t) = [n_1(t)\, n_2(t)\, \ldots\, n_M(t)]^T$$

and $\boldsymbol{A}(\theta)$ is M×D matrix of the steering vectors:

$$\boldsymbol{A}(\theta) = [\boldsymbol{a}(\theta_1)\, \boldsymbol{a}(\theta_2)\, \ldots\, \boldsymbol{a}(\theta_D)] \qquad (1)$$

$s_i(t)$ – the signal of the $i^{th}$ source
$n_i(t)$ – the noise at $i^{th}$ sensor
The steering vector of the array towards direction $\theta$ is
$$\boldsymbol{a}(\theta) = \left[e^{j\frac{2\pi}{\lambda}(x_1\cos(\theta)+y_1\sin(\theta))} \ldots e^{j\frac{2\pi}{\lambda}(x_M\cos(\theta)+y_M\sin(\theta))}\right]^T$$
where $(x_i, y_i)$ is the coordinate of the $i^{th}$ sensor, $j = \sqrt{-1}$ and $\lambda$ is wavelength of the signal impinging on the array.

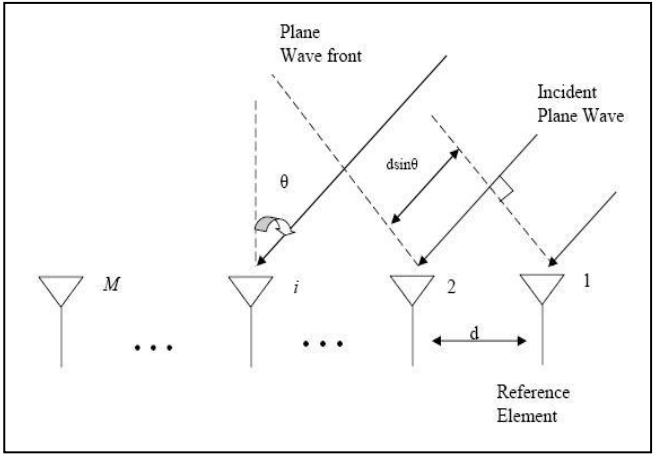

FIG 1 Array signal model for uniformly spaced sensors

The DOA estimation problem is to estimate the locations of the source (angles $\theta_1, \theta_2 \ldots \theta_D$) from the N samples (snapshots) of the array $\boldsymbol{u}(t_1), \boldsymbol{u}(t_2) \cdots \boldsymbol{u}(t_N)$.

### III. PROPOSED ALGORITHMS

#### A. Root extension to Propagator Algorithm

For the case of a uniformly spaced linear array with inter-element spacing 'd' the $m^{th}$ element of the steering vector $\boldsymbol{a}(\theta)$ may be expressed as below
$$a_m(\theta) = \exp\left[j \cdot 2\pi \cdot m \left(\frac{d}{\lambda}\right) \cdot \cos\theta\right] \quad m = 1,2 \ldots M \quad (2)$$
Without any loss of generality, assumption is made that the array is located along the x-axis, hence eq. (2) from eq. (1).

Under the hypothesis that the steering matrix $\boldsymbol{A}$ is of full rank, D rows of $\boldsymbol{A}$ are linearly independent. The other rows of $\boldsymbol{A}$ can be expressed as a linear combination of these D rows. Without loss of generality, hereafter it is assumed that the first D rows are linearly independent.

Thus, the propagator is based on the partition of the steering vector $\boldsymbol{A}$ from equation (1) according to:
$$\boldsymbol{A} = \begin{bmatrix} \boldsymbol{A_1} \\ \boldsymbol{A_2} \end{bmatrix} \begin{matrix} \} \ D \\ \} \ M-D \end{matrix}$$
where $\boldsymbol{A_1}$ and $\boldsymbol{A_2}$ are matrices of dimensions D x D and (M - D) x D respectively.

Definition of Propagator: Under the hypothesis that $\boldsymbol{A_1}$ is non-singular, propagator is the unique linear operator $\boldsymbol{P}$ mapping $\mathbb{C}^{M-D}$ into $\mathbb{C}^D$, equivalently defined as follows:
$$\boldsymbol{P}^H \cdot \boldsymbol{A_1} = \boldsymbol{A_2}$$
where $\boldsymbol{P}$ is D x (M-D) matrix. Or it can be written as

$$\boldsymbol{P}^H \cdot \boldsymbol{A_1} - \boldsymbol{I_{M-D}} \cdot \boldsymbol{A_2} = \boldsymbol{0}$$
$$[\boldsymbol{P}^H, -\boldsymbol{I_{M-D}}] \begin{bmatrix} \boldsymbol{A_1} \\ \boldsymbol{A_2} \end{bmatrix} \triangleq \boldsymbol{Q}^H \boldsymbol{A} = \boldsymbol{0} \quad (3)$$
where $\boldsymbol{I_{M-D}}$ and $\boldsymbol{0}$ are the identity matrix and the null matrix of dimension (M – D) and (M – D) x D, respectively.

Thus, from equation (3), Q is defined as follows
$$\boldsymbol{Q} = [\boldsymbol{P}^H, -\boldsymbol{I_{M-D}}]^H \quad (4)$$
Now, the DOA estimator is defined as the Propagator spectrum ($\boldsymbol{F_{PM}}$), an all pole function of the form
$$\boldsymbol{F_{PM}}(\theta) = \{\boldsymbol{a}^H(\theta) \boldsymbol{Q} \boldsymbol{Q}^H \boldsymbol{a}(\theta)\}^{-1} \quad (5)$$
$$\boldsymbol{F_{PM}}(\theta) = \{\boldsymbol{a}^H(\theta) \boldsymbol{C} \boldsymbol{a}(\theta)\}^{-1}$$
where $\boldsymbol{C} = \boldsymbol{Q}\boldsymbol{Q}^H$

Using the above equation the denominator can be written as
$$\boldsymbol{F_{PM}}^{-1}(\theta) = \sum_{m=1}^{M}\sum_{n=1}^{M}\left(e^{\left[-j\frac{2\pi d}{\lambda}m\cos\theta\right]}C_{mn}e^{\left[j\frac{2\pi d}{\lambda}n\cos\theta\right]}\right)$$
where $C_{mn}$ is the entry in the $m^{th}$ row and $n^{th}$ column of $\boldsymbol{C}$. Combining the two summations into one, above equation can be simplified as given below – equation (6)
$$\boldsymbol{F_{PM}}^{-1}(\theta) = \sum_{l=-M+1}^{M-1}\left(C_l \cdot e^{\left[-j\frac{2\pi d}{\lambda}l\cos\theta\right]}\right)$$
where
$$C_l \triangleq \sum_{l=m-n} C_{mn}$$
is the sum of the entries of $\boldsymbol{C}$ along the $l^{th}$ diagonal. The polynomial $\boldsymbol{D}(z)$ is defined as given below, equation (7)
$$\boldsymbol{D}(z) \triangleq \sum_{l=-M+1}^{M-1} C_l * z^{-l}$$
Evaluating the Propagator spectrum $\boldsymbol{F_{PM}}(\theta)$ is equivalent to evaluating the polynomial $\boldsymbol{D}(z)$ on the unit circle. The peaks of the Propagator spectrum exist as roots of $\boldsymbol{D}(z)$ lying close to the unit circle. Ideally, with no noise, the poles will lie exactly on the unit circle at locations determined by the DOA [27]. In other words, the pole of $\boldsymbol{D}(z)$ at
$$z = z_1 = |z_1| \cdot \exp(j * arg(z_1))$$
will result in the peak of the Propagator spectrum at
$$\cos\theta = \left[\frac{\lambda}{2\pi d} * arg(z_1)\right] \quad (8)$$
Simulations show that Root-Propagator achieves superior resolution in comparison with the Propagator algorithm, especially at lower SNR conditions. And simulations also show that at high SNR, Root-Propagator and Propagator have almost the same RMSE values. At low SNR, the Root-Propagator algorithm provides a better resolving capacity over Propagator algorithm, but with degradation in the RMSE performance. Thus, a modification is proposed to the Root-Propagator algorithm for further improvement of its performance.

#### B. Advanced Root-Propagator algorithm

First the DOA of the signals are found out using the Root-Propagator algorithm as given by eq. (8) and sorted.

Now, a threshold is defined such that it minimizes the unresolved failures at the least computation cost by limiting the search for DOA in the Propagator Spectrum.

Next for each of these angles, the propagator spectrum, given by eq. (5), is scanned simultaneously to both the left and the right starting from the angle under consideration in incremental steps (0.01° here) until a peak is encountered or till the threshold is reached, whichever is earlier.

This process is repeated for each of these angles found using Root-Propagator method.

If a peak is encountered in the Propagator spectrum within the defined threshold, then it is taken as the estimated DOA else the angle estimated by the Root-Propagator algorithm is taken as the estimated angle.

If any two DOAs estimated in Propagator spectrum are same i.e. exactly overlapping with the output of the Root-Propagator, then the DOA estimated by the Root-Propagator algorithm is considered as the estimated angle.

Thus, the number of unresolved failures in the algorithm is reduced, enabling an improved resolution capacity without loss in RMSE.

## IV. SIMULATION AND OUTPUTS

The important parameters for a DOA estimation algorithm are RMSE, angular resolution and complexity.

Root Mean Square Error (RMSE) is a direct measure of accuracy, a parameter that defines how precisely the given algorithm can estimate the angle of arrival (AOA) of the input signals. RMSE is calculated using the equation (9)

$$RMSE = \sqrt{\frac{1}{L} * \sum_{i=1}^{L}(\theta_i - \hat{\theta}_i)^2}$$

where, $\hat{\theta}_i$ is the estimate of actual DOA $\theta_i$ and L is the total number of trials. The given algorithm is considered to have successfully resolved if the difference in the estimated angle and the actual angle-of-arrival is less than the given threshold value which is taken as 7°.

Angular resolution is defined as the least angular spacing between two signals-of-interest (SOI) of equal signal strength, which can be clearly distinguished.

A comparison of RMSE, angular resolution and simulation time for all the algorithms – Propagator, Root-Propagator and Advanced Root-Propagator algorithms are performed in this section.

The following parameters were used for all the simulation results presented in this paper: Number of source D = 2, Number of Sensors M = 12, Number of Snapshots N = 200, center frequency $\omega_c$ = 850 MHz, Resolution of 0.01 degrees Threshold value = 7 degrees and Number of trials L = 200.

The RMSE performance and the angular resolution for all the algorithms are simulated using MATLAB®. The simulation considered here are the following cases for SNR -10 dB to 10 dB in steps of 5 dB for two sets of angles first (40, 50) degrees and second (62, 70) degrees.

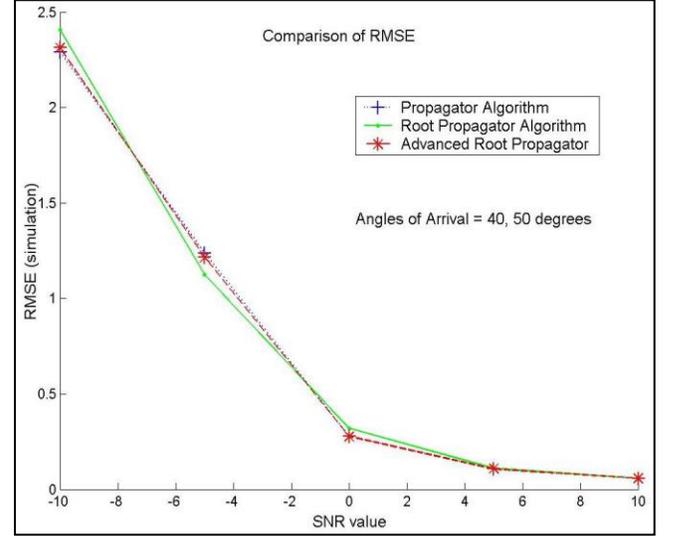

FIG 2: Comparison of RMSE for all algorithms for 40° & 50° angles of arrival. This shows the new algorithms perform well.

In fig. 2 and fig. 3 the RMSE values of Propagator, Root-Propagator and Advanced Root-Propagator have been plotted. From these graphs, one can conclude that at low SNRs, the Advanced Root-Propagator algorithm has almost the same RMSE value as that of the Propagator algorithm, but the Root-Propagator algorithm has higher RMSE performance. However, at high SNRs, all the three algorithms have similar RMSE performance.

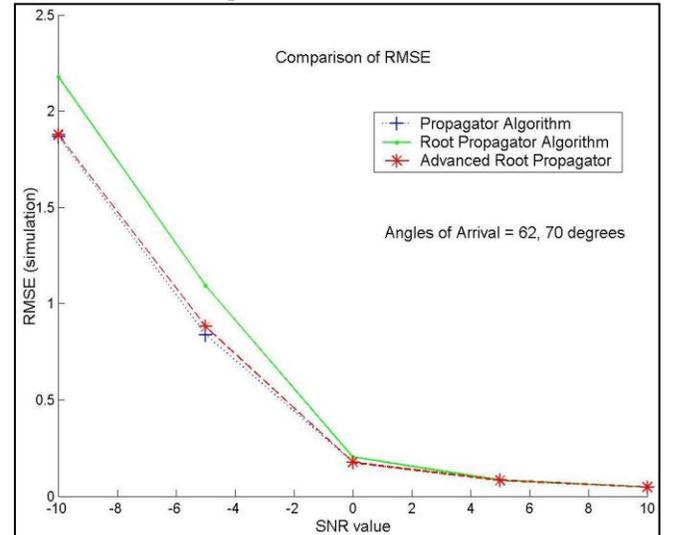

FIG 3: Comparison of RMSE for all algorithms for 62° & 70° angles of arrival. This shows the proposed algorithms perform equally well.

In fig. 4, the number of unresolved failures has been plotted for the Advanced Root-Propagator algorithm considering different thresholds and for various SNR values. From this graph, one can conclude that the number of unresolved failures initially decreases with the increase in threshold value but becomes constant above a certain threshold value which is 5º in this case. One can also conclude that the threshold value for the Advanced Root-Propagator algorithm decreases with the increase in the SNR and the threshold value for -10dB SNR are around 5º.

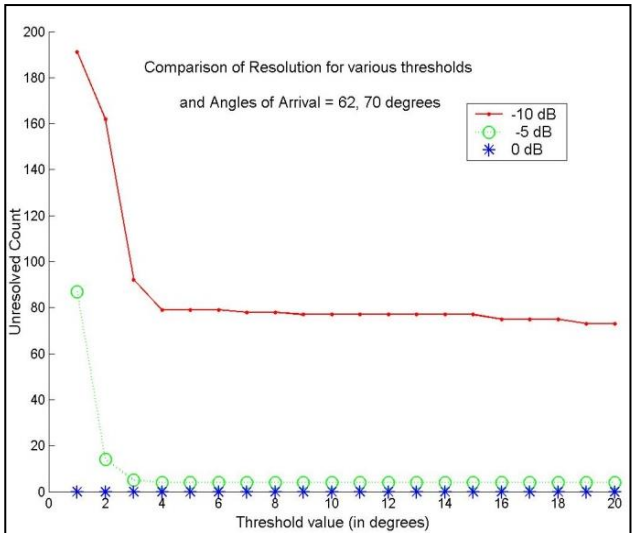

FIG 4: Comparison of resolution for various thresholds for 62º & 70º angles of arrival. This shows that the threshold value to be used in the advanced Root Propagator algorithm increases with decreasing SNR.

In Table I, a comparison of the number of unresolved failures (or angular resolution) has been made for the various algorithms for various SNR. These show that the angular resolution of the Root-Propagator is better than that of the Propagator algorithm. Also, the angular resolution of the Advanced Root-Propagator algorithm is better or almost equal to that of the Root-Propagator algorithm.

TABLE I. UNRESOLVED FAILURES FOR ALL ALGORITHMS AT 40 AND 50 DEGREES (G1) AND 62 AND 70 DEGREES (G2) AT DIFFERENT SNR VALUES. PROPOSED ALGORITHMS REDUCES FAILURES SIGNIFICANTLY COMPARED TO THE PROPAGATOR METHOD

| Algorithm (for 200 trials) | | SNR values (dB) | | | | |
|---|---|---|---|---|---|---|
| | | -10 dB | -5 dB | 0 dB | 5 dB | 10 dB |
| Propagator | G1 | 189 | 50 | 0 | 0 | 0 |
| | G2 | 127 | 40 | 0 | 0 | 0 |
| Root-Propagator | G1 | 135 | 1 | 0 | 0 | 0 |
| | G2 | 86 | 4 | 0 | 0 | 0 |
| Advanced Root-Propagator | G1 | 134 | 1 | 0 | 0 | 0 |
| | G2 | 48 | 4 | 0 | 0 | 0 |

## V. COMPARISON OF COMPUTATIONAL COMPLEXITY

A comparison of complexity of the different DOA estimating algorithms discussed in this paper is presented in this section viz. MUSIC, Root MUSIC, Propagator Method, and the proposed algorithms in Table II. There are a few points that one should note. Propagator method achieves a complexity reduction over MUSIC of the order of the ratio of number sources to number of sensors used by avoiding EVD. Root Propagator method further achieves complexity reduction by avoiding spatial spectrum search for peaks. The slight complexity increase in advanced root propagator over Root Propagator is due to spectrum search and is marginal. There are a variety of polynomial rooting algorithms with cubic and lower complexity [28, 29] hence the orders of complexity due to polynomial rooting is not explicitly shown.

TABLE II. COMPARISON OF THE THEORETICAL COMPUTATIONAL COMPLEXITY FOR THE DOA ESTIMATION ALGORITHMS DISCUSSED

| Algorithm | Computational Complexity |
|---|---|
| MUSIC | $O(M^3 + P \times M \times D)$ |
| Root-MUSIC | $O(M^3 + degree\ M\ rooting)$ |
| Propagator | $O(M^2 D + P \times M \times D)$ |
| Root-Propagator | $O(M^2 D + degree\ M\ rooting)$ |
| Advanced Root-Propagator | $O(M^2 D + degree\ M\ rooting + Q \times M \times D)$ |

where M is the number of sensors; D is the number of DOAs to be estimated. P is the number of angles at which we evaluate the spectrum for MUSIC and propagator algorithms. It is defined as the ratio of the total angle to be scanned, in this case entire range i.e. 180º to the resolution of the scan (0.01º in this case). While Q is the number of angles at which we evaluate the spectrum for advanced root-propagator algorithm. It is the ratio of the total angle to be scanned, in this case it is threshold (i.e. 5º) to the resolution of the scan (0.01º in this case).

Simulation time is a good indicator of the computation complexity of the algorithm. In Table III, the average simulation time of various algorithms over 16,000 trials has been tabulated. This table shows that the simulation time required for the Advanced Root-Propagator algorithm is slightly higher than that of the Root-Propagator algorithm due to partial search in spatial spectrum yet it is much lesser than that of the Propagator algorithm.

TABLE III. COMPARISON OF AVERAGE SIMULATION TIME REQUIRED PER ITERATION FOR VARIOUS ALGORITHMS. PROPOSED ALGORITHMS REDUCES COMPLEXITY SIGNIFICANTLY COMPARED TO THE PROPAGATOR

| Algorithm | Avg. Simulation Time (ms) |
|---|---|
| Propagator method | 1,808.19 |
| Root-Propagator | 16.13 |
| Advanced Root-Propagator | 24.44 |

The proposed algorithms achieve around 98% reduction in complexity when compared over the traditional Propagator algorithm for estimating the DOA with comparable RMSE performance.

## VI. Conclusion

We show that the proposed Root-Propagator algorithms have better angular resolution, lesser RMSE even at lower SNR while being computationally less complex by a huge factor when compared over the traditional Propagator method. Thus, the proposed algorithms Root-Propagator and Advanced Root-Propagator algorithm efficiently combine the advantages of both the Propagator and the Root-MUSIC algorithms. These improvements are beneficial in the Internet of Things domain where large number of resource constrained devices are used which necessitate the deployment of power efficient implementations. However, since the proposed algorithms assume uniformly spaced linear array of sensors, future work should address the need to improve these methods for non-uniform sensor spacing and non-linear arrays as well as 2D array of sensors.